# Capacity Analysis in Multi-Radio Multi-Channel Cognitive Radio Networks: A Small World Perspective

Xiaoxiong Zhong, Yang Qin, Li Li

**Abstract** Cognitive radio (CR) has emerged as a promising technology to improve spectrum utilization. Capacity analysis is very useful in investigating the ultimate performance limits for wireless networks. Meanwhile, with increasing potential future applications for the CR systems, it is necessary to explore the limitations on their capacity in a dynamic spectrum access environment. However, due to spectrum sharing in cognitive radio networks (CRNs), the capacity of the secondary network (SRN) is much more difficult to analyze than that of traditional wireless networks. To overcome this difficulty, in this paper we introduce a novel solution based on small world model to analyze the capacity of SRN. First, we propose a new method of shortcut creation for CRNs, which is based on connectivity ratio. Also, a new channel assignment algorithm is proposed, which jointly considers the available time and transmission time of the channels. And then, we derive the capacity of SRN based on the small world model over multi-radio multi-channel (MRMC) environment. The simulation results show that our proposed scheme can obtain a higher capacity and smaller latency compared with traditional schemes in MRMC CRNs.
**Keywords** Capacity Analysis. Cognitive Radio Networks. Small World

## 1 Introduction

The CR principle has introduced the idea to exploit spectrum holes (i.e., bands) which result from the proven underutilization of the electromagnetic spectrum by modern wireless communication and broadcasting technologies [1]. The exploitation of these holes can be accomplished by the notion of cognitive radio networks (CRNs). CRNs have emerged as a prominent solution to improve the efficiency of spectrum utilization and network capacity. In CRNs, secondary users (SUs) can exploit frequency bands when the primary users (PUs) do not occupy the bands. The objective of CRNs is to optimize the performance, e.g., the capacity of the SRN, without causing harmful effects on PUs. Existing research works on capacity of CRNs have mainly focused on improving the performance of the physical layer and media access control (MAC) [2]. These approaches can provide high capacity in single-hop topology, which are ineffective in multi-hop scenarios. For example, an optimized sensing threshold method may provide a higher capacity for a particular link. However, such a method may be inefficient when considering the average path length of a given multi-hop CRN.

Capacity analysis is very useful in investigating the ultimate performance limits for CR systems. Some efforts have been taken to improve the CR channel capacity through optimizing the lower-layer parameters [2-3]. In a CRN, the power control and the spectrum sensing are properly incorporated for the capacity optimization providing the PU's protection [4]. In [5], Jararian *et al.* studied a symmetric multiuser cognitive radio system and presented a lattice coding scheme which all the *L* transmit-receive pairs can simultaneously communicate as if all cross channels were absent from the system. In [6], assuming a path loss shadow-fading model with multiple PUs and SUs, the system-level capacity of CRNs under an average interference power constraint has been investigated. Their results show that the uplink ergodic channel capacity of a CR-based central access network can be relatively large when the number of PUs is small. In [7], Li *et al.* analyzed the achievable throughput by using the characteristics of the single hop transport throughput of the SRN with outage constraints. In [8], capacity and delay scaling laws are introduced for cognitive radio ad hoc networks by capturing the impact of PU activity in dense and sparse PU deployment conditions.

The above mentioned proposals of capacity analysis are restricted to the link level, where only one CR transmitter opportunistically communicates with one CR receiver in the presence of a single or multiple primary receivers. They mainly focus on how to deal with the interference channel setting or power control. Taking end-to-end path into account, it is shown that cross layer design can maximize the sum commodity throughput through jointly optimize transmission power, constellation size, temporal schedule, and the corresponding optimal flow [9]. In [10], it is demonstrated that joint

School of Computer Science and Technology, Harbin Institute of Technology Shenzhen Graduate School, Shenzhen, 518055, People's Republic of China.
Xiaoxiong Zhong, e-mail: xixzhong@gmail.com
Yang Qin, e-mail: yqinsg@gmail.com
Li Li, e-mail:lili8503@aliyun.com

optimization at physical (power control), link (scheduling), and network (flow routing) layers with the signal-to-interference-plus-noise ratio (SINR) function will increase the capacity of the network. In these schemes, average path length is a key parameter for capacity analysis from a network layer perspective. However, the capacity optimization of SRN considering the average path length has not been studied even though it is more important to obtain higher achievable capacity in CRNs.

Small world is a fundamental and important topological property of complex networks [12], existing in many reality networks such as Internet, product relations network and electric network [13]. It is characterized by the facts that clustering coefficient is large and average path length is short. In 2003, Helmy first introduced the concept of the small world into wireless networks and proved that small world phenomena also exist in wireless networks [14]. Guidoni *et al.* proposed a small world model to design MRMC heterogeneous sensor networks, which takes the communication pattern of this network to create shortcuts directed to the monitoring node [15]. The simulation results show that it can reduce the latency significantly and the number of collisions during the data dissemination. However, it does not consider the heterogeneous availability of channels in CRNs. Also, they didn't consider the capacity of SRN from the end-to-end path perspective. Due to the heterogeneity in channel availability over CRNs, the capacity is much more difficult to characterize. Therefore, we should take the connectivity ratio between any two nodes into account, which based on channel availability, for capacity analysis in CRNs.

In 2013, Azimdoost *et al.* exploited the social behavior to analyze the capacity of wireless networks, modeling the user behavior based on probabilistic information [16]. Different from their work, in our scheme, we use queuing theory and small world model to analyze the capacity of a given SRN. In addition, we consider the channel availability in capacity analysis.

To the best of our knowledge, using small world to analyze the CRNs performance has not been theoretically studied. In this work, we exploit the small world concept to analyze the capacity of a given SRN. First, spectrum opportunity of licensed channels is modeled by preemptive resume priority queue model; also we propose a novel scheme to create shortcuts and a new channel assignment algorithm in CRNs. Subsequently, the capacity of SRN, which includes single radio and multiple radios scenarios, is derived. The main contributions of this paper are four-fold: 1) give a novel method to calculate the connectivity ratio in MRMC CRNs, which is a key parameter for capacity analysis; 2) propose a scheme for shortcut creation based on the connectivity ratio in CRNs; 3) a new channel assignment algorithm is proposed, which jointly considers the available time and transmission time of the channels; and 4) the capacity of SRN is derived based on the small world model over MRMC CRNs.

The rest of this paper is structured as follows. The system model including spectrum opportunity modeling, shortcut creation in a small world and channel assignment is described in Section 2. We derive the capacity of SRN in MRMC scenarios in Section 3. We undertake a thorough performance evaluation in Section 4. Section 5 concludes the paper.

## 2  System Model

### A.  *Spectrum Opportunity (SOP) Modeling*

In our previous work [17], we assume an interweave model [18], i.e., the nodes in the CRNs can only transmit when PUs are not active. Thus, there is virtually no interference to the PUs and there is no restriction on the power level of the SUs. We also assume that the set of SOP is the same for every node in the same primary networks.

We consider the CRNs where PUs and SUs access the network through the shared channels. The total channels of the network contains *N* channels based indexed by *n* ($n=1, 2,…, N$).

*Primary network* (PRN): there are *N* users (PUs) in primary network, and the *N* orthogonal channels are assigned to the *N* Pus, respectively. It means that each PU could only occupy one licensed channel.

*Secondary network* (SRN): there are *M* users (SUs) in our secondary network.

In CRNs, each channel has two types of customers (PU and SU). The PUs have the preemptive priority to interrupt the transmission of the SUs. The remaining transmission of the interrupted SU will be put into the head of the low priority queue of the current operating channel. Furthermore, the interrupted SU can resume the unfinished transmission when the current channel becomes idle, instead of retransmitting the whole data. A SU transmission may experience multiple interruptions from the PUs during the data transmission period. Thus, we model the cognitive network system using a preemptive-resume priority queuing model [19], as shown in Fig.1. In our model, a virtual queue with different priorities is utilized to model the traffic of PUs and SUs on the same channel. Utilization of a channel by PUs can be modeled by an M/M/1/1 queue [19]. Utilization of a channel by SUs can be modeled by an M/M/N/K queue [19]. In the model, we assume that the arrival processes of the PUs and the SUs are Poisson. Let $\lambda_p$ and $\lambda_s$ be the average arrival rates of the primary connections on the channel and the secondary connections, respectively. Each PU is further characterized by transmissions with average duration of $1/\mu_p$, (PU-service rate, $\mu_p$) and the average transmission duration time of a SU is $1/\mu_s$, (SU-service rate, $\mu_s$). The service policy of each queue is first in first out (FIFO).

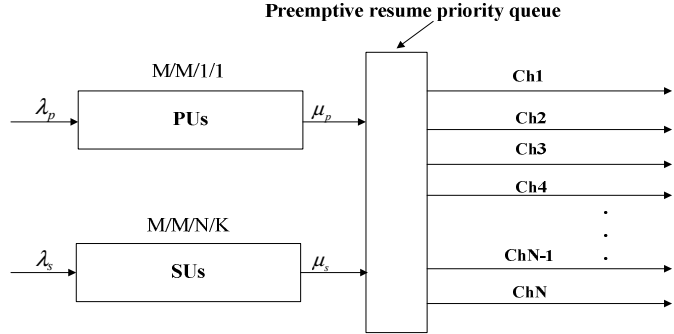

Fig. 1. Preemptive resume priority queue for licensed channels.

According the M/M/1/1 queue model, we get the balance condition

$$\begin{cases} \lambda_p p_{p0} = \mu_p p_{p1} \\ p_{p0} + p_{p1} = 1 \end{cases} \quad (1)$$

where $p_{p0}$ is the probability that the channel is in idle state (denoted as state 0), $p_{p1}$ is the probability that the channel is occupied by the PU (denoted as state 1). In order to make the queue to be stable, we have to require $\rho_p < 1$, where $\rho_p = \lambda_p / \mu_p$.

Thus, the probability of channel that the PU does not occupy the channel is

$$p_{p0} = \frac{1}{1+\rho_p} = \frac{1}{1+\lambda_p/\mu_p} \quad (2)$$

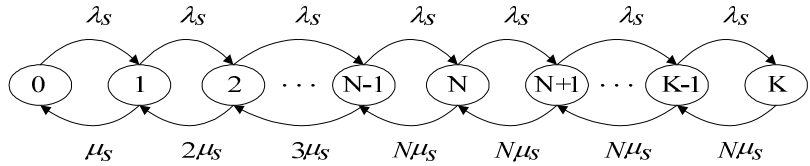

Fig. 2. Flow diagram for the M/M/N/K model for SUs.

The queue model of SUs is shown in Fig.2, according to the M/M/N/K queue model, we get the stationary distribution, which can be written as

$$p_{si} = \begin{cases} \dfrac{N^i (\rho_s)^i}{i!} p_{s0} & 0 \leq i < N \\ \dfrac{N^N (\rho_s)^i}{N!} p_{s0} & N \leq i \leq K \end{cases} \quad (3)$$

where $\rho_s = \lambda_s / N\mu_s$.

Due to the normalizing condition, we can obtain the probability $p_{s0}$ that all SUs do not occupy the channel when PU is inactive on the channel.

$$p_{s0} = \left[ \sum_{k=0}^{N-1} \frac{(N\rho_s)^k}{k!} + \frac{(N\rho_s)^N}{N!} \frac{1-\rho_s^{K-N+1}}{1-\rho_s} \right]^{-1} \quad (4)$$

The probability that there are $h$ channels occupied by SUs in all $N$ licensed channels.

$$p_{sh} = \begin{cases} \dfrac{N^h(\rho_s)^h}{h!} p_{s0}, (0 \le h < N) \\ \dfrac{N^N(\rho_s)^h}{N!} p_{s0}, (N \le h \le M) \end{cases} \quad (5)$$

And, we can obtain the probability that the $K^{th}$ channel occupied by a SU in $h$ channels

$$p_{sK} = \dfrac{p_{sh}}{\binom{N}{h}} \quad (6)$$

The conditional probability that there are exactly $h$-$1$ channels occupied by the other $h$-$1$ SUs under the scenario that the $K^{th}$ channel is occupied by a SU.

$$p_{sh-1/h} = \binom{N-1}{h-1} p_{sK} = \dfrac{h}{N} p_{sK} = \begin{cases} \dfrac{h}{N} \dfrac{N^h \rho_s^h}{h!} p_{s0}, (0 \le h < N) \\ \dfrac{h}{N} \dfrac{N^N \rho_s^h}{N!} p_{s0}, (N \le h \le M) \end{cases} \quad (7)$$

Hence, we can obtain the probability that the $k^{th}$ channel is occupied by a SU is

$$p_{ij}^k = p_{sh-1/h} p_{p0} = \begin{cases} p_{p0} \dfrac{h}{N} \dfrac{N^h \rho_s^h}{h!} p_{s0}, (0 \le h < N) \\ p_{p0} \dfrac{h}{N} \dfrac{N^N \rho_s^h}{N!} p_{s0}, (N \le h \le M) \end{cases} \quad (8)$$

We note that if $i = j$, $p_{ij}^k = 1$.
Thus, we define the connectivity ratio between any two nodes $i, j$ as

$$p_{ij} = 1 - (1 - p_{ij}^1)(1 - p_{ij}^2)...(1 - p_{ij}^N) \quad (9)$$

**B. *Shortcut Creation***

Small world is captured by two measurements: small average path length and high clustering coefficient (defined as the average fraction of pairs of neighbors of a node that are also neighbors of each other). Using small world concept could optimize the performance of wireless networks through creating shortcuts [14-15], which will reduce the average hop count of the network. In [14], a shortcut is defined as a logical link that a node maintains with a randomly selected node. Therefore, a logical link between a pair of nodes can eventually correspond to several physical hops, and this logical link is implemented by making each node in the path to keep a route to the long-range contact. In [15], Guidoni *et al.* proposed a new method to create shortcuts directed to the monitoring node for MRMC heterogeneous sensor networks. However, they do not explore the heterogeneous availability of channels in any two nodes. Also, in [15], it needs more time to search the shortcut-candidate. For a CRN, it is inefficient to select the shortcut-candidate because of the uncertainty of channel, which is subjected to PU activity. Therefore, it is necessary to develop a new method to find the shortcut-candidates for CRNs. To address this problem, we propose a new shortcut creation method (NSC) based on connectivity ratio, which is structured as follows. Note that in our scheme, hubs have to be equipped with more than two radios, one to communication with the hub, and the others to communication with normal SUs. Fig. 3 illustrates the process of shortcut creation.

**Step 1:** Selecting the nodes that have a high degree as hubs over CRNs. If a node and its neighbors allow to forward many packets, its degree in the graph is high. As a result, a node $i$ decides to be a hub when it has relatively large degree locally and the *available channel set* of the node is not null.

**Step 2:** Determining $i$'s searching region for creating shortcut. Node $i$ sends a *hello* message that contains its ID, geographic position and *available channel set* using the specific radio. Also, each hub stores the received *hello* messages in a neighbor table. It calculates the straight line $l_{iD}$ that passes the geographic position of $i$ and destination node ($D$). This line is the bisectrix of an angle $α$. Select the node $C$ as a candidate, which has the highest degree in $i$'s neighbor table. The straight line $l_{CD}$ passes the geographic position of *C* and *D*. The straight line $l_{iC}$ passes the geographic position of *i* and *C*. The angle $\theta_1$ is formed by line $l_{iD}$ and $l_{iC}$. The angle $\theta_2$ is formed by line $l_{iD}$ and $l_{CD}$. Hence, the *i*'s searching region is the closed region formed by the three straight lines $l_{iD}$, $l_{CD}$ and $l_{iC}$.

**Step 3:** Selecting shortcut-candidates from the searching region, we choose the shortcut-candidates for a given hub using by following inequalities

$$\tan\theta_1 < \tan(\alpha/2) \quad (10)$$

$$\tan\theta_2 < \tan(\alpha/2) \quad (11)$$

If the inequalities are satisfied, we choose the nodes which are inside the searching region as the *i*'s shortcut-candidates. Then, in *i*'s shortcut-candidates, we select the node that has maximal connectivity ratio $p_{ij}$ (*j* is *i*'s shortcut candidate) between node *i* and node *j*. Also, the node *j* should have a high degree.

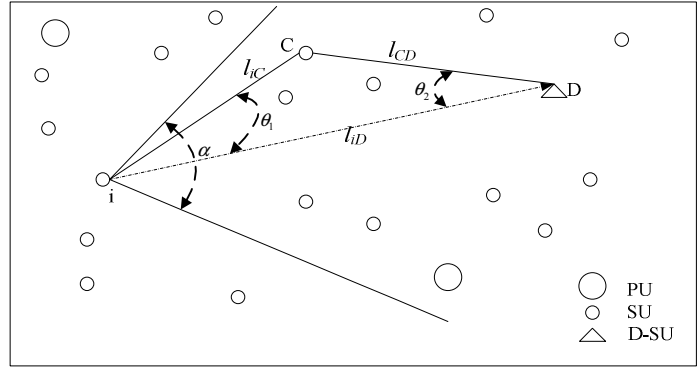

Fig. 3. Shortcut creation method.

After executing all above steps, each hub has a candidate for creating shortcut. It sends a *create-shortcut* message to the candidate. If the candidate has not created a shortcut yet, it sends an *ack* message, creating shortcut. If not, it sends a *nack* message. In this case, the hub will choose a different candidate that currently has the highest connectivity ratio in its shortcut-candidates and again sends a *create-shortcut* message. If all candidates of a hub return *nack* messages, the hub randomly chooses a neighbor that the *available channel set* of it is null and sends a *force-create-shortcut* message.

## C. Channel Assignment

Table 1. Summary of key notations

| Symbol | Meaning |
|---|---|
| *i* | The node in CRNs which runs **Algorithm 1** |
| *Neighbor* | Set of *i*'s neighbors |
| *Avai* | Set of node *i*'s available channels |
| *AvaiN* | Set of neighboring nodes' *Avai* |
| *R* | Set of radios on node *i* |
| *X* | Set of channels in *Avai* which are not assigned |
| *Z* | Set of radios in *R* on which channel is not assigned |
| *C(R)* | Set of channels tuned to set of radios *R* |
| $T_r$ | Transmission time of a given channel |
| $T_v$ | Available time of a given channel |
| *Y* | Set of hubs in CRNs |

After shortcut creation over MRMC CRNs, we must choose and allocate the channel from the available channel set to the radio of a shortcut. Channel assignment can either be centralized [20] [21] [22] or distributed [23] [24] over CRNs. A centralized approach to the channel assignment usually obtains best results. However, it typically brings about a high communication overhead. In addition, the channel availability of CRNs is frequently time-varying; the centralized policies become less efficient. Thus, we design a distributed approach in our solution, such as [25]. However, our scheme jointly considers the transmission time and available time of an available channel, which is shown in

**Algorithm 1**. We summarize some notations that we use in the **Algorithm 1**, as shown in Table 1.

According to our work [30], we can obtain the transmission time $T_r$ and the available time $T_v$ of a given channel. Each node will assign the first channel (that has minimal $\Delta$) to its first radio and the second channel to its second radio and so on. We assume that if a channel is assigned, it is immediately used to transmit data packet. Note that if a PU arrives on the assigned channel then only the affected nodes will update the *available channel set* and repeat the **Algorithm 1**.

---

**Algorithm 1** Channel assignment algorithm

1: **Input:** *i*, *Neighbor*, *Avai*, *AvaiN*, *R*, *Y*
2: $Z \leftarrow R$, $X \leftarrow Avai$, $Temp \leftarrow Avai \cap AvaiN$
3: **while** $Temp \neq \emptyset$ **do**
4:    Sort *Temp* by value of $\Delta = T_v - T_r$ in ascending order
5:    Update *Temp* with removing the channels that have negative value of $\Delta$
6:    **while** $Z \neq \emptyset$ **do**
7:       If $i \in Y$ then
8:          Select *i'* shortcut-candidate and create a shortcut between them using special radio
9:          Select a channel *c* that has minimal $\Delta$
10:         $C(R) \leftarrow c$, $Z \leftarrow Z/\{z\}$, $X \leftarrow X/\{c\}$
11:      **else**
12:         Select a node from *i'* *Neighbor*
13:         Select a channel *c* that has minimal $\Delta$
14:         $C(R) \leftarrow c$, $Z \leftarrow Z/\{z\}$, $X \leftarrow X/\{c\}$
15:      **end if**
16:   **end while**
17: **end while**
18: **Output** $C(R)$

---

## 3 Capacity Analysis in CRNs

Reducing the average path length will improve the capacity of the network when taking end-to-end path into account. As mentioned earlier, using the small world concept can significantly reduce average path length of the networks. The challenge in capacity analysis of CRNs has motivated us to design a small world model with the MRMC paradigm for secondary system of CRNs.

Fig. 4 shows an example of CRNs in which consists of primary network (PRN) and the secondary network (SRN). The PUs and SUs are distributed in the same area and they share the frequency bands (channels). The black nodes are PUs and the white nodes are SUs. The small dotted circle denotes SU's transmission range and the big dotted circle represents PU's transmission range.

**Theorem 1.** The network capacity (the maximum value of data transmission in a CRN) of SRN over MRMC CRNs based on small world mode, is

$$Capa = M * T' * F * (1 - \frac{\tau}{T_s}) * [(1-P_f)P(H_0) + (1-P_d)P(H_1)]$$

and the effective capacity that the maximum value of sending data from the source node to the destination node, is

$$Capa_e = \frac{M * T'}{L(G)} * F * (1 - \frac{\tau}{T_s}) * [(1-P_f)P(H_0) + (1-P_d)P(H_1)]$$

where *M* is the number of SUs in CRNs, $T' = T_0 / [p_{ij}^{average}((4-2/k) + C(G)(k-1)(1-1/k))]$ is the effective capacity per node, $T_0$ is the available capacity of a SU, *k* is the average degree of node, *F* is a factor causing by MRMC, $L(G)$ is average path length, $C(G)$ is clustering coefficient, $P_f$ is false alarm probability, $P_d$ is detection probability, $P(H_0)$ is the PU-free probability, $P(H_1)$ is PU-busy probability, $\tau$ is sensing time, $T_s$ is time slot, and $p_{ij}^{average}$ is average connectivity ratio between any two node *i, j* in a SRN.

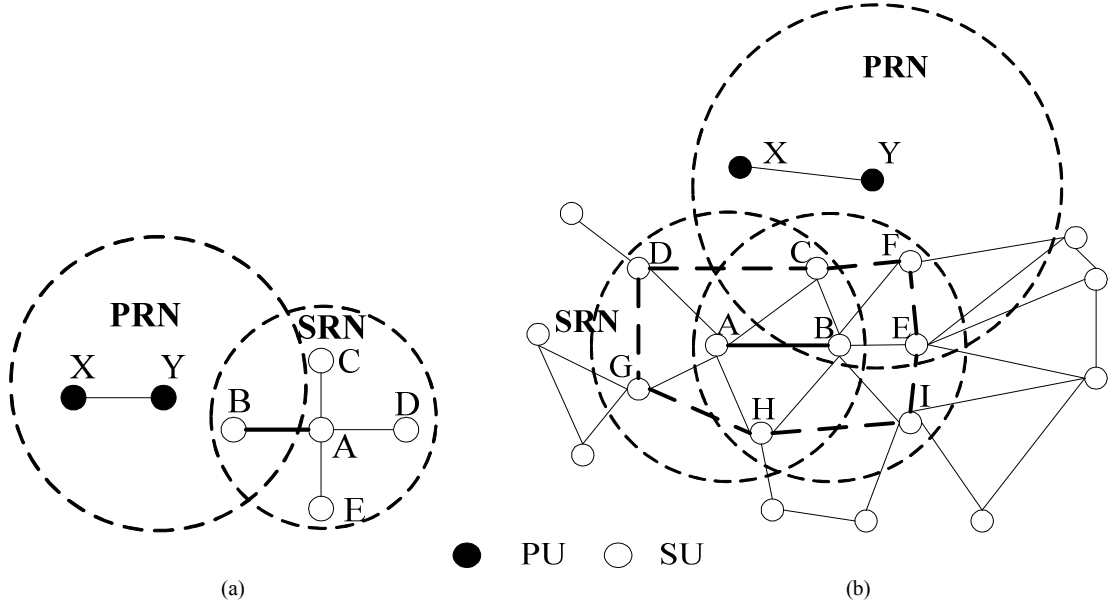

Fig. 4. An example of CRNs coexisting of two different systems (PRN and SRN).

**Proof:** In [26], Gupta and Kumar have shown that the capacity of a wireless mesh network diminishes as the number of nodes increasing in the network. It is not to be expected that the available throughput to each node approaches zero as the number of nodes increases. In the following analysis, we first derive the capacity of SRN in single channel single radio scenario, and then give the capacity in MRMC scenario.

Assume the maximum per-node capacity in CRNs is $T_0$, for a given SRN with $M$ nodes, the maximal achievable capacity to all nodes is $M*T_0$. However, in single channel single radio CRNs, the MAC is adopted CSMA/CA, when a node transmits data, the data transmissions of its neighbors are prohibited. Thus, when a link is used to transmit data, it consumes network resource (in this paper, we only consider nodes and links). Generally, in each data transmission, it consumes more than two nodes. Next, we discuss how many nodes are consumed in each data transmission.

In Fig. 4(a), when node $A$ sends data to $B$, the neighbors of $A$ are restricted to transmit data, thus they cannot transmit data to any node. In this case, this transmission will consume all the neighbors of node $A$, i.e., nodes $C$, $D$, and node $E$.

In Fig. 4(b), when node $A$ sends data to $B$, their neighbors cannot transmit data to any node in the transmission range, i.e., nodes $C$, $D$, $E$, $F$, $G$, $H$, and node $I$. The link $BC$ is totally in the intersection of $A$ and $B$ transmission range. In this case, node $C$ is totally consumed for $A{\rightarrow}B$ transmission. Link $FE$ and link $IE$ are in transmission range of node $B$, whereas they are out of the intersection of the transmission range of node $A$ and node $B$. It is referred as a partial consumption of node $A$ and node $B$. Thus, the consumption ratio of node $E$ consumed by nodes $A$ and $B$ is $(1+1*\gamma+1*\beta)/k_E$, where $\gamma/\beta$ is the consumption ratio of node $E$ by node $A/B$, and $k_E$ is the degree of node $E$. Similarly, when $B$ transmits data to $A$, we can obtain the consumption ratio.

In consumed-link set, it includes the total consumption, e.g., link $BC$ and the partial consumption, e.g., link $FE$. In the following, we give the method to calculate the number of the consumed links, taking the $A{\rightarrow}B$ transmission for an example.

The total consumption $f_1$ is the sum of the degree of nodes $A$ and $B$, subtracting by double counting, which is expressed as

$$f_1 = 2k_A + 2k_B - 2 \qquad (12)$$

where $k_A$ and $k_B$ are the degree of nodes $A$ and $B$, respectively.

The partial consumption $f_2$ is the sum of the adjacent edges of the neighbors of nodes $A$ and $B$.

The clustering coefficient $C$ is a measure of the degree to which nodes in a graph tend to cluster together. $C_A$ and $C_B$ are the clustering coefficient for nodes $A$ and $B$, respectively. $f_A$ and $f_B$ are the partial consumption of nodes $A$ and $B$. Thus, we can obtain

$$f_A = C_B k_B (k_B - 1) \qquad (13)$$

$$f_B = C_A k_A (k_A - 1) \tag{14}$$

Assume the role of a node in the SRN is identical. Thus, $f_A$ ($f_B$) can be rewritten as

$$f_A = C_B k_B (k_B - 1)(1 - 1/k_B) \tag{15}$$

$$f_B = C_A k_A (k_A - 1)(1 - 1/k_A) \tag{16}$$

Thus, the partial consumption is

$$f_2 = [C_B k_B (k_B - 1)(1 - 1/k_B)] + [C_A k_A (k_A - 1)(1 - 1/k_A)] \tag{17}$$

Therefore, when node $A$ sends data to node $B$, the consumed links $f$ is

$$f = 2k_A + 2k_B - 2 + \beta [C_B k_B (k_B - 1)(1 - 1/k_B)] + \gamma [C_A k_A (k_A - 1)(1 - 1/k_A)] \tag{18}$$

In the SRN, the probability $\gamma$ ($\beta$) that node $E$ is in transmission range of node $A$ ($B$) is very large. Thus, in our analysis, we assume that $\gamma$ is 0.5 and $\beta$ is 0.5. In this case, $f$ can be rewritten as

$$f = 2k_A + 2k_B - 2 + 0.5[C_B k_B (k_B - 1)(1 - 1/k_B)] + 0.5[C_A k_A (k_A - 1)(1 - 1/k_A)] \tag{19}$$

If we consider node $A$ and node $B$ to be identical nodes in the SRN, we can replace $k_A$ ($k_B$) and $C_A$ ($C_B$) with mean degree $k$ and mean clustering coefficient $C(G)$, respectively.

As pointed out in [27], PU's activity is one of the dominant features that affect the secondary spectrum access. Thus, it is necessary to consider PU's activity for analyzing the capacity of secondary system in CRNs. The mean connectivity ratio, $p_{ij}^{average}$, between any two node $i, j$ in the SRN is

$$p_{ij}^{average} = \frac{\sum_{i=1}^{M} \sum_{j=1}^{M} \sum_{k=1}^{N} p_{ij}^k}{\frac{NM(M-1)}{2}} \quad (i \neq j) \tag{20}$$

where $N$ is the number of channels and $M$ is the number of SUs in the SRN.

Thus we can obtain the total consumed links in a data transmission:

$$f = p_{ij}^{average} (4k - 2 + C(G)k(k-1)(1 - 1/k)) \tag{21}$$

And, the average value of the consumed nodes in a data transmission is

$$k_0 = p_{ij}^{average} [(4 - 2/k) + C(G)(k-1)(1 - 1/k)] \tag{22}$$

Hence, the effective capacity per node is

$$T' = T_0 / [p_{ij}^{average} ((4 - 2/k) + C(G)(k-1)(1 - 1/k))] \tag{23}$$

Finally, we can obtain the capacity,

$$Capa = M * T' \tag{24}$$

and the effective capacity,

$$Capa_e = M * T' / L(G) \tag{25}$$

The SUs are allowed to opportunistically utilize spectrum bands assigned to PUs as long as they are inactive. Thus, the SUs need to periodically sense the presence of PUs to utilize their occupied frequency bands. In this communication paradigm, the probabilities of detection and false alarm are especially important. Therefore, in our analysis, we should consider detection probability, false alarm probability and the PU-free probability [28]. The capacity and effective capacity of the SRN can be represented as,

$$Capa = M*T'*(1-\frac{\tau}{T_s})*[(1-P_f)P(H_0)+(1-P_d)P(H_1)] \tag{26}$$

$$Capa_e = \frac{M*T'}{L(G)}*(1-\frac{\tau}{T_s})*[(1-P_f)P(H_0)+(1-P_d)P(H_1)] \tag{27}$$

where $P_f$ is false alarm probability, $P_d$ is detection probability, $P(H_0)$ is PU-free probability, $P(H_1)$ is PU-busy probability, $\tau$ is sensing time, and $T_s$ is the time slot.

In MRMC scenario, the node with multi-radio may simultaneously transmit data. Ideally, the capacity is $M*T_0$ and effective capacity is $M*T_0 / L(G)$. Due to the limited channel and radio, it makes sense to use an enhancement factor $F$ to characterize the capacity improvement based on MRMC paradigm. Thus, we set $F$ according to [29]. In [29], we know that the capacity of multi-channel arbitrary networks is limited the number of channels $N$ and the number of radios $R$, where the upper bound is $O(W\sqrt{\frac{MR}{N}})$ under interference constraint, where $M$ is the number of nodes and $W$ is the fixed data rate of each channel. Therefore, we have

$$Capa = M*T'*F*(1-\frac{\tau}{T_s})*[(1-P_f)P(H_0)+(1-P_d)P(H_1)] \tag{28}$$

$$Capa_e = \frac{M*T'}{L(G)}*F*(1-\frac{\tau}{T_s})*[(1-P_f)P(H_0)+(1-P_d)P(H_1)] \tag{29}$$

From the discussion above, it is observed that for a given value of $M$ and $T_0$, the capacity can be increased by maximizing the ratio of

$$\frac{M*T_0}{[(4-\frac{2}{k})+C(G)(k-1)(1-\frac{1}{k})]*p_{ij}^{average}*L(G)} \tag{30}$$

In this case, increasing the capacity of each node is likely to give rise to greater radio interference over a larger coverage area due to the need for a higher transmit power. Alternatively, effective capacity can be increased by reducing the average path length $L(G)$, which in turn represents a decrease in number of hops traversed by a packet. The latter approach is adopted in this paper as it will not only increase throughput but also minimize transmission delay.

## 4 Numerical and Simulation Results

In this section, we present simulation results to evaluate the effective capacity and data dissemination latency efficiency benefits of the proposed scheme. We conduct the simulations with varying sensing time, the number of shortcuts creation, and channel availability over these schemes: without small world (Without SW), traditional small world with randomly selecting node to create shortcut [14], (RS), and our new shortcut creation scheme (NSC) under the scenarios: random channel assignment (Random) [31] and the proposed channel assignment (CA), respectively. Also, we compare NSC with DM-MC [15] in terms of data dissemination latency.

We set up a CRN with 12 PUs and 100 SUs randomly distributed in a 1000×1000 m² area. The network parameter settings are shown in Table 2. In addition, PU-service rate $\mu_p$ and SU-service rate $\mu_s$ are randomly chosen from the interval [0, 1]. The channel availability is 0.8 (except this scenario: capacity vs. channel availability, as shown in Fig.9). The sensing time $\tau$ is 10ms (except this scenario: capacity vs. sensing time, as shown in Fig.7). The values Latency (0) and L(0) represent the data dissemination latency and average path length, respectively, when the CRN does not have shortcuts. Values Latency (N) and L(N) represent the same metrics when the CRN has shortcuts. The ratio between L(N)/L(0) and Latency(N)/Latency(0) is calculated for following results.

Fig. 5 shows the data dissemination latency of DM-MC scheme and our proposed scheme, NSC under different shortcuts creation scenario. When the number of shortcut increases, the latency reduces. On average, the latency of the NSC is 25% smaller than that of DM-MC. This is because the searching region of DM-MC is larger than NSC during shortcut creation, which means that it needs more time to create shortcuts in DM-MC. In addition, we take the connectivity ratio into account for shortcut creation in NSC, which can reduce the PU-SU collision probability for data transmission. Hence, the data dissemination latency of NSC is smaller.

Table 2 Simulation parameters

| | |
|---|---|
| Number of channels | 12 |
| Number of radios | 4 |
| Number of PUs | 12 |
| Number of SUs | 100 |
| PU transmission range | 100m |
| SU transmission range | 50m |
| Capacity available of a SU $T_0$ | 2Mbps |
| Average degree of the network $k$ | 4 |
| Average clustering coefficient of the network $C(G)$ | 0.4 |
| Average arrival rates of PU $\lambda_p$ | 0.2 |
| Average arrival rates of SU $\lambda_s$ | 0.5 |
| Bisectrix of an angle $\alpha$ | 30 degrees |
| Detection probability $P_d$ | 0.9 |
| False alarm probability $P_f$ | 0.2 |
| PU-free probability $P(H_0)$ | 0.5 |
| Time slot $T_s$ | 100ms |
| Packet size | 512 bytes |

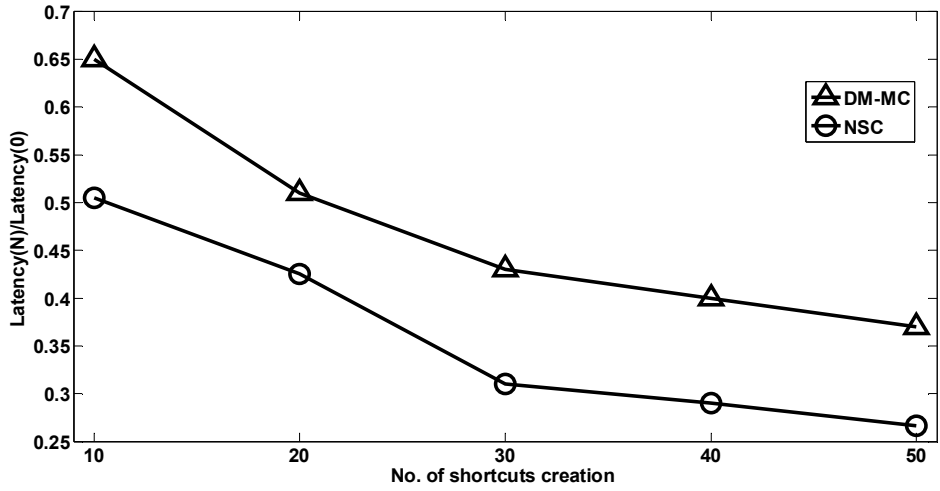

Fig. 5. Latency vs. No. of shortcuts creation.

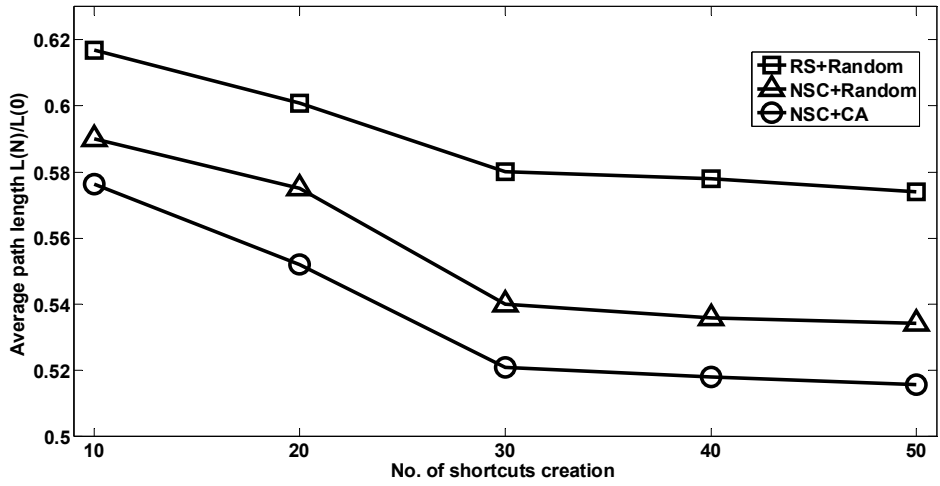

Fig. 6. Average path length vs. No. of shortcuts creation.

In Fig. 6, we can see that the average path length reduces as shortcuts increase. When the number of shortcuts is smaller than 30, the CRN does not present small world feature. However, when it is larger than 30, the average path length does not reduce significantly over these three schemes. In this case, it is no necessary for more than 30 shortcuts to deploying in the CRN. The NSC+CA presents smaller values of average path length than RS+Random and NSC+Random, which means that it is easy to find a route with lower average path length in the CRN with NSC+CA scheme. This is because it has smaller searching the region for shortcut creation the considering connectivity ratio in NSC+Random, while jointly considering channel availability and channel available time in NSC+CA.

In Fig. 7, we compare the following four schemes: Without SW, RS+Random, NSC+Random and NSC+CA, in terms of capacity for different sensing time. We can see that NSC+CA performs better than other three schemes, which jointly considers the connectivity ratio in shortcut creation and channel available time in channel assignment. The total capacity of these schemes decreases as the sensing time increases. Because of increasing sensing time, the time for transmitting data will be reduced proportionally. Also, the results imply that it can reduce the average hop count of the network through creating shortcuts, resulting in increasing the effective capacity of networks.

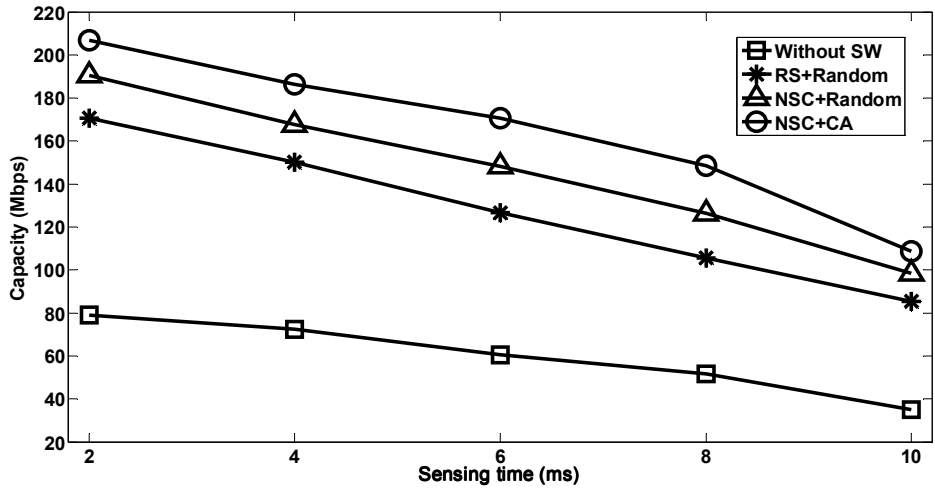

Fig. 7. Capacity vs. Sensing time (10 shortcuts).

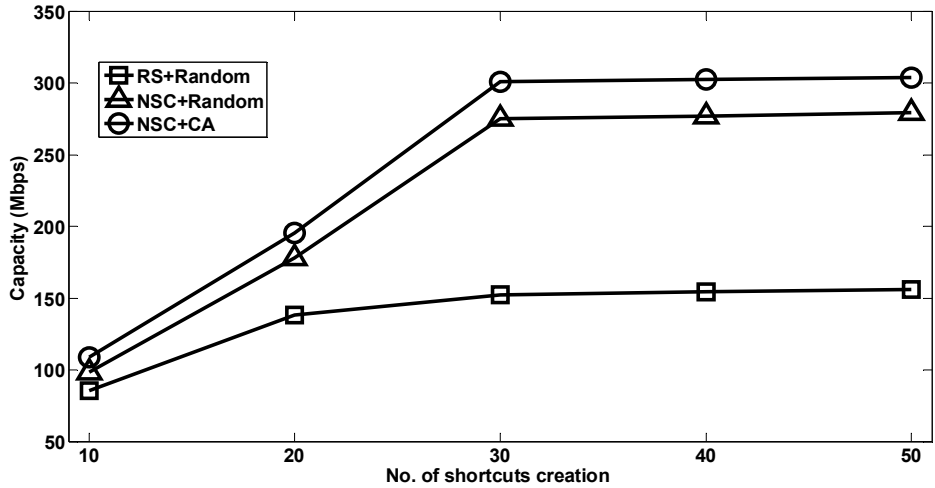

Fig. 8. Capacity vs. No. of shortcuts creation.

In Fig. 8, with increasing the number of shortcuts, the capacity of RS+Random, NSC+Random and NSC+CA schemes increase as well. The NSC+CA achieves higher capacity than other two schemes, which mainly considers the connectivity ratio in shortcut creation and channel available time in channel assignment. It will reduce the PU-SU collision probability for data transmission and be more quickly to create shortcuts. Hence, in NSC+CA, it has a shorter average path length, resulting in improving the capacity of the networks. In the meantime, when the number of shortcuts is more than 30, the effective capacity does not increase significantly. This is because the average path length does not reduce significantly over these three schemes.

In Fig. 9, we illustrate the capacity of Without SW, RS+Random, NSC+Random and NSC+CA for different channel availability, when the number of shortcuts is 30. When the channel availability increases, the capacity increases as well. This is because when the channel availability is very small, there will be not enough available channels for data transmission, when the channel availability is high, it is easy to create shortcuts and find a route with lower average path length to transmit the data. In addition, NSC+CA is effective in fully utilizing the channel, which takes connectivity ratio and the channel available time into account. This in turn affects the average path length. Hence, the NSC+CA achieves higher capacity than other two schemes. Also, the simulation results show that NSC is better than RS and the proposed channel assignment algorithm CA performs better than Random. In addition, the results imply that using small world can reduce the average hop count of the network, resulting in increasing the effective capacity of networks.

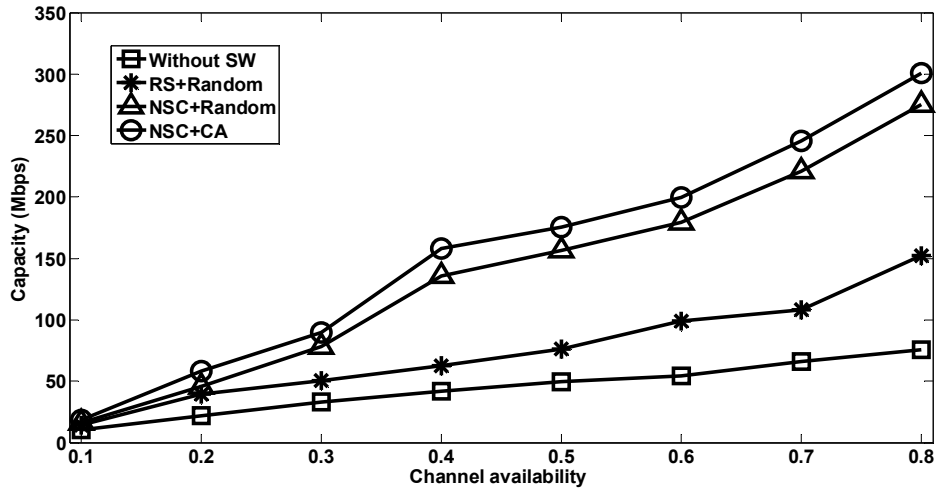

Fig. 9. Capacity vs. Channel availability.

## 5 Conclusions

At present, capacity analysis of CRNs mainly focuses on link layer via setting interference channel or operating power control algorithms. Taking end-to-end path into account to analyze the capacity of CRNs is still an open issue. The average path length is shorter in the small world model through adding shortcuts. In this paper, we propose a novel shortcut creation method based on connectivity ratio, which has a lower latency compared with DM-MC. In the meantime, a new channel assignment algorithm is proposed, which considers the available time and transmission time of the channels. And then, the capacity analysis has been proposed and proved theoretically over MRMC CRNs. By dynamically adding shortcuts on the small world model over MRMC CRNs, the system capacity has increased significantly compared with the traditional schemes. In our scheme, both data dissemination latency and system capacity have been improved significantly and verified by numerous simulation results.

Our further studies on capacity analysis in multi-hop CRNs need to pay much attention on topology control and appropriate power control scheme for SUs and PUs.

**Acknowledgements** Financial supports from the Shenzhen Science and Technology Fundament Research Foundation (No. JC200903120189A, JC201005260183A, and ZYA201106070013A) are highly appreciated.